\documentclass{webofc}
\usepackage[varg]{txfonts}   
\newcommand{\apj}{Astrophys. J. }%
\newcommand{\aap}{Astron. Astrophys. }%
\newcommand{\aj}{Astron. J. }%
\newcommand{\nar}{New Astron. Rev. }%
\newcommand{\nat}{Nature }%
\newcommand{\physrep}{Phys.~Rep. }%
\newcommand{\Al}{$^{26}$Al\ }

\newcommand{\Msol}{M\ensuremath{_\odot}}

\newcommand{\revised}[1]{{ #1}}
\begin{document}
\title{Gamma-ray spectroscopy of galactic nucleosynthesis}
%
%
\author{\firstname{Roland} \lastname{Diehl}\inst{1}\fnsep\thanks{\email{rod@mpe.mpg.de}} 
	\and
        \firstname{Jochen} \lastname{Greiner}\inst{1}
              \and
        \firstname{Martin G.H.} \lastname{Krause}\inst{2}
        \and
        \firstname{Moritz M.M.} \lastname{Pleintinger}\inst{1,3}
        \and
        \firstname{Thomas} \lastname{Siegert}\inst{4}
}
   \institute{Max-Planck-Institut f\"ur extraterrestrische Physik, Giessenbachstr. 1, D-85748 Garching, Germany
   \and
      Centre for Astrophysics Research, School of Physics, Astronomy and Mathematics, University of Hertfordshire, College Lane, Hatfield, Hertfordshire AL10 9AB, UK
  \and
   Horn \& Company Financial Services GmbH, Kaistr. 20, D-40221 D\"usseldorf, Germany
	   \and
	Institut f\"ur Theoretische Physik und Astrophysik, Universit\"at W\"urzburg, Emil-Fischer-Str. 31, D-97074 W\"urzburg, Germany
          }
\abstract{%
 Diffuse gamma-ray emission from the decay of radioactive $^{26}$Al  is a messenger \revised{from} the nucleosynthesis activity in our current-day galaxy. Because this material is attributed to ejections from massive stars and their supernovae, the gamma-ray signal \revised{includes information} about \revised{ nucleosynthesis in} massive star \revised{interiors as it varies with evolutionary stages, and about} their feedback \revised{on the}  surrounding interstellar medium. Our method of population synthesis of massive-star groups has been refined as a diagnostic tool for this purpose. It allows to build a bottom-up prediction of the diffuse gamma-ray sky when known massive star group distributions and theoretical models of stellar evolution and core-collapse supernova explosions are employed. We find general consistency of an origin in such massive-star groups, in particular we also find support for the clumpy distribution of such source regions across the Galaxy, and characteristics of large cavities around these. A discrepancy in the integrated $^{26}$Al gamma-ray flux is interpreted as an indication for excess $^{26}$Al emission from nearby, distributed in cavities that extend over major regions of the sky. 
}
\maketitle
\section{Introduction}
\label{sec:intro}
  $^{26}$Al has been established as a tracer of ongoing nucleosynthesis in our Galaxy \cite{Prantzos:1996a}. 
Spectroscopy at high resolution was essential for the pioneering discovery, because it provided an unambiguous line identification at 1809~keV as known from the laboratory. COMPTEL only featured scintillation detectors and hence a modest spectral resolution of $\sim$200~keV at this energy, unlike Ge detectors, where 2-3~keV resolution can be achieved. 
 With INTEGRAL \cite{Winkler:2003} and its SPI gamma-ray spectrometer \cite{Vedrenne:2003,Roques:2003}, also featuring Ge detectors, a new step in spectroscopic detail of $^{26}$Al emission could be achieved, as summarised in \cite{Diehl:2021}.
   The INTEGRAL mission has enabled accumulation of observing time since its 2002 launch, continuing to this day, measuring the large-scale Galactic $^{26}$Al emission. 
In this paper we \revised{summarize} results from analysis of 17.5 years of data, using both single and double detector events within the SPI camera  \revised{for the first time (see} \cite{Pleintinger:2020,Pleintinger:2023} \revised{for more detail)}, and drawing on a significantly-enhanced version of our population synthesis code \cite{Voss:2009} \revised{as developed in  \cite{Pleintinger:2020}} for interpretations of our results.

\section{Data and analysis}
\label{sec:data}
 INTEGRAL operates since October 2002 in an excentric orbit located outside the radiation belts.
The SPI spectrometer is one of the two main instruments on INTEGRAL, specialised for high-resolution spectroscopy over the
15~keV to 8~MeV energy range with a  camera \revised{composed or 19 Ge detectors}, achieving an energy resolution of 3~keV at 1809~keV. 
A coded mask above the camera shadows some of the detectors from emission of each region of the sky, while others obtain exposure of these regions. 
\revised{The  instruments aboard the INTEGRAL satellite are co-aligned, and have rather large fields of view, SPI having an effective opening angle of $\sim$30~degrees. Observations are carried out by pointing the common instrument axes towards a region of interest on the sky. Then,}   
offsetting the instrument pointings by 2.1 degrees approximately every 30 minutes in a regular rectangular pattern \revised{(this is called \emph{dithering})} provides \revised{additional detector shadowing variations} that \revised{ help  to} effectively \revised{achieve} an imaging resolution of about 2.6 degrees at 1809~keV. 
For individual source regions in the plane of the Galaxy, a sensitivity (3$\sigma$) of 2.5$\times$10$^{-5}$~ph~cm$^{-2}$s$^{-1}$
has been achieved by now (from a 6~Ms exposure in the example of the Vela region) \cite{Pleintinger:2020}.

Here we use data from 2002 until 2020, i.e., 17.5 years of observations, with 2131 INTEGRAL orbits of typically 3 days duration each.
After filtering out periods of abnormal conditions, 118407 spacecraft pointings from 1840 orbits remain, providing a sky exposure of about 255~Ms.
For each of these pointings, the data from single Ge detector hits, as well as those where two Ge detectors triggered within their 350~ns coincidence window (called \emph{double events, DE}) were binned into energy spectra for each of the \revised{19} detectors named 00-18; pairs of DE detectors are assigned detector ID's above 18, i.e., 19-60. Detectors are numerated from central towards outside in left-spiral counting, so ID=0 is the central detector and detectors 07-18 are on the outside of the hexagonal dense pack forming SPI's camera.
In total, during the mission four of the nineteen Ge detectors failed (ID's 01,02,05,17), \revised{at times} between orbit numbers 140 and 930. \revised{ The} camera response \revised{differs} for different numbers of operational detectors (\revised{from 19 down to 15}), in particular the rates of single versus double or multiple detector triggers \revised{changes} (because failed detectors do not provide a 'multiple-event' trigger any more), \revised{which needs to be accounted for in the analysis}. 

Event data from each detector are calibrated using known background lines, so that the distortions through gain variations are corrected for. Then these events are binned into spectra at 0.5~keV width, covering the 20--2000~keV range for response and background, and the 1790--1840~keV range for spectroscopy of the \Al line.
At the $^{26}$Al line energy of 1809 keV, the number of double-hit events corresponds to 56\% of the amount of single-hit events, all including instrumental background and the celestial $^{26}$Al signal.

These SPI detector count spectra  are  
$d_{i,j,k}$ measured counts, with $i,j,k$ being the indices of pointing, detector, and energy bin.  
The 19 physical detectors are supplemented by the respective detector combinations for multiple-detector hits, resulting in a detector variable $j\in[0,60]$ instead of $j\in[0,18]$.
We describe these data as the result of the instrument's response to the $\gamma$-ray sky, plus an underlying instrumental background:
\begin{equation}
d_{i,j,k} = \sum_l R_{l;ijk} \sum_{n = 1}^{N_\mathrm{s}} \theta_n S_{nl} +
\sum_{n = N_\mathrm{s} + 1}^{N_\mathrm{s} + N_\mathrm{b}} \theta_n
B_{n;ijk}\label{eq:model-fit}
\end{equation}
\revised{Herein, $\theta_n$ and $\theta_b$ are model parameters scaling sky and background models, $S$ or $B$, respectively.} Sky models $S_{nl}$ are formulated as photon source intensities per sky direction $l$, \revised{and multiple source or spectral components $n$ can be used for sky and for background}. For best sensitivity at high spectral resolution, we avoid simultaneous imaging, given our large number of free parameters of background modelling to ensure its closest match to real background. 
The instrument response matrix $R_{l;ijk}$ implements the \revised{shadowgram} coding by mask and dithering, thus linking the emission \revised{of a pixel $l$ in the sky}  to \revised{pixels in data space}.
The instrumental background \revised{$B$} is modelled from detailed spectral fits to raw data. \revised{Integrating} over suitable intervals \revised{results in smearing out} any celestial contributions, while still retaining \revised{sufficient} temporal resolution  to trace temporal variations of background (see \cite{Diehl:2018} and \cite{Siegert:2019} for detailed descriptions of this background and its modelling).
\revised{Maximum-likelihood fits then optimise model parameters through a } comparison of the data as measured to predictions from these models of sky and background, thus obtaining the energy spectrum \revised{$\theta_n$} of intensities for the sky model, as a result.
	
\revised{Among the wide range of plausible sky models  (see \cite{Prantzos:1996a,Diehl:1995i,Knodlseder:1999a} , for a discussion), we used} the image obtained from the COMPTEL 9-year allsky survey \cite{Diehl:1995b,Pluschke:2001c} as our \revised{baseline} sky model.
Then we aim \revised{for} astrophysical implications \revised{of our data, formulating} models from astrophysical considerations of the spatial distribution of \Al sources in our Galaxy. \revised{We have been} using tracers that plausibly may represent these \Al sources , \revised{such as maps of molecular gas or emission from warm dust, or distributions of free electrons attributed to massive-star ionizations} \cite{Diehl:1997,Knodlseder:1999a}.
We now advanced this approach to construct a full bottom-up model of  \revised{the number and spatial distribution of the} \Al sources \revised{in} our  Galaxy \cite{Pleintinger:2020}.
For this, we start from models of stars with their evolution and nucleosynthesis yields, building stellar groups using a mass spectrum, as we had done in our earlier population synthesis approach \cite{Voss:2009}. Now we place these groups into a spatial model for the Galaxy, using astronomical knowledge about massive star groups where accessible  \revised{(i.e., catalogues derived from a variety of types of observations; these are available} mainly in the vicinity of the solar system to a distance of few kpc at most), \revised{and we supplement these  for the more-distant part of the Galaxy with} Monte-Carlo samples \revised{of analytical models for the} large-scale distribution, \revised{such as exponentially shaped disks or spiral arm models. We approximate this} 
 large-scale \revised{source} distribution either through a Gaussian \revised{function based on} the pulsar distribution \cite{Yusifov:2004}, or an exponential profile with its characteristic scale radius. The vertical structure above the Galactic plane is modelled as an exponential. 
We superimpose spiral structure herein, as derived from observables such as the pulsar emission dispersion measure, converted into the electron distribution in the Galaxy \cite{Cordes:2002}.  For our description, we use a parametrisation of the four spiral arms following \cite{Faucher-Giguere:2006}. 
The main astrophysical parameters of this bottom-up modelling, which we aim to determine comparing its predictions to our data, are:
(i) The absolute intensity of Galactic recent massive-star activity. This is implemented in a \emph{star formation rate},  and scaled until the model flux matches the all-sky flux observed in the \Al line.
(ii) The scale radius and width of the large-scale distribution. We decided to use five representative cases: Spiral arm models with a truncated gaussian \revised{(using} widths 1.8, 2.5, and 3.5~kpc, and scale radii of 8.5, 17.5, and 5~kpc, respectively, for \revised{one} four-arm and two two-arm variants\revised{)}, and an exponential disk \revised{(}with a fixed scale radius of 5.5~kpc following the pulsar distribution but without spiral structure\revised{), plus} a generic exponential centrally peaked in the Galaxy. 
(iii) The vertical \revised{ extent of \Al emission, i.e. its} scale height.
Drawing random realisations of such a model by sampling the parameters included in such a bottom-up stars-groups-galaxy model, one obtains a stellar population synthesis based model of a galaxy. We call this latest approach the \emph{PSYCO} implementation \cite{Pleintinger:2020,Siegert:2023}.
We supplemented this generic and randomly-drawn model with a less-randomized treatment of massive star clusters where we have additional knowledge, either from calalogues, or from deeper studies including our own \Al measurements. For the few clusters where we were able to derive a gamma-ray brightness in $^{26}$Al, we used these as brightness normalisations, rather than drawing randomly, thus without a need to draw a sample age and richness in stars. For the ones where only a position was known, we used this position. We started from these nearby clusters with some knowledge, and then continued sampling randomly from the spatial distribution model until the integrated model flux agreed with the measured flux in our dataset; this determines the normalization of our large-scale model, i.e. the corresponding star formation rate.
This provides us with a predictive model for the appearance of the observed \Al sky, evaluating the Galaxy's content of \Al from theory inputs concerning source evolution and nucleosynthesis output.
We then varied our model parameters, and determined a best-fit statistics based on likelihood with Poissonian statistics, to select best parameters for our measured dataset.
	

\section{Results}
\label{sec:results}

\begin{figure}[ht] 
\centering
\includegraphics[width=0.32\columnwidth,clip]{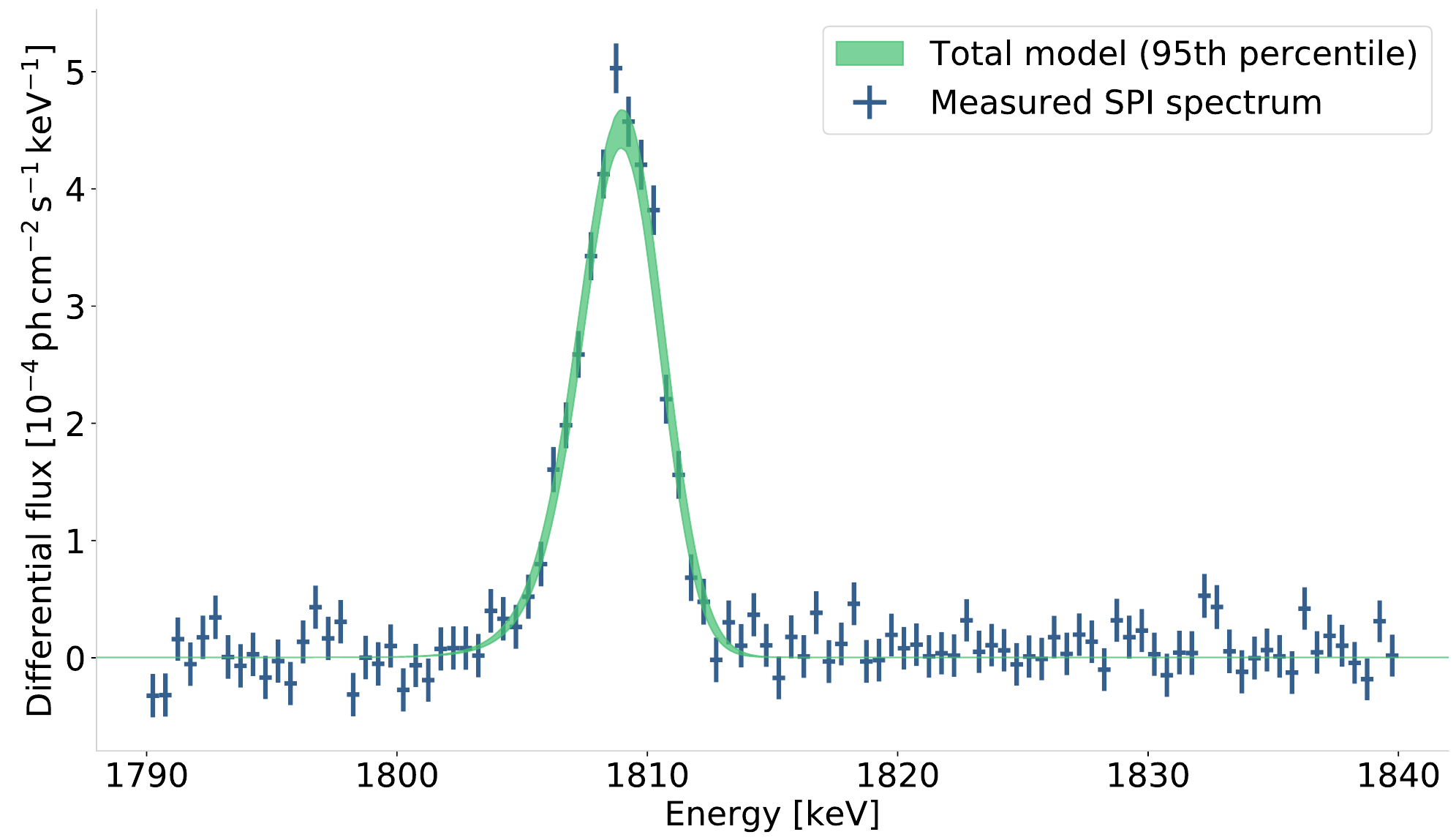}
\includegraphics[width=0.32\columnwidth,clip]{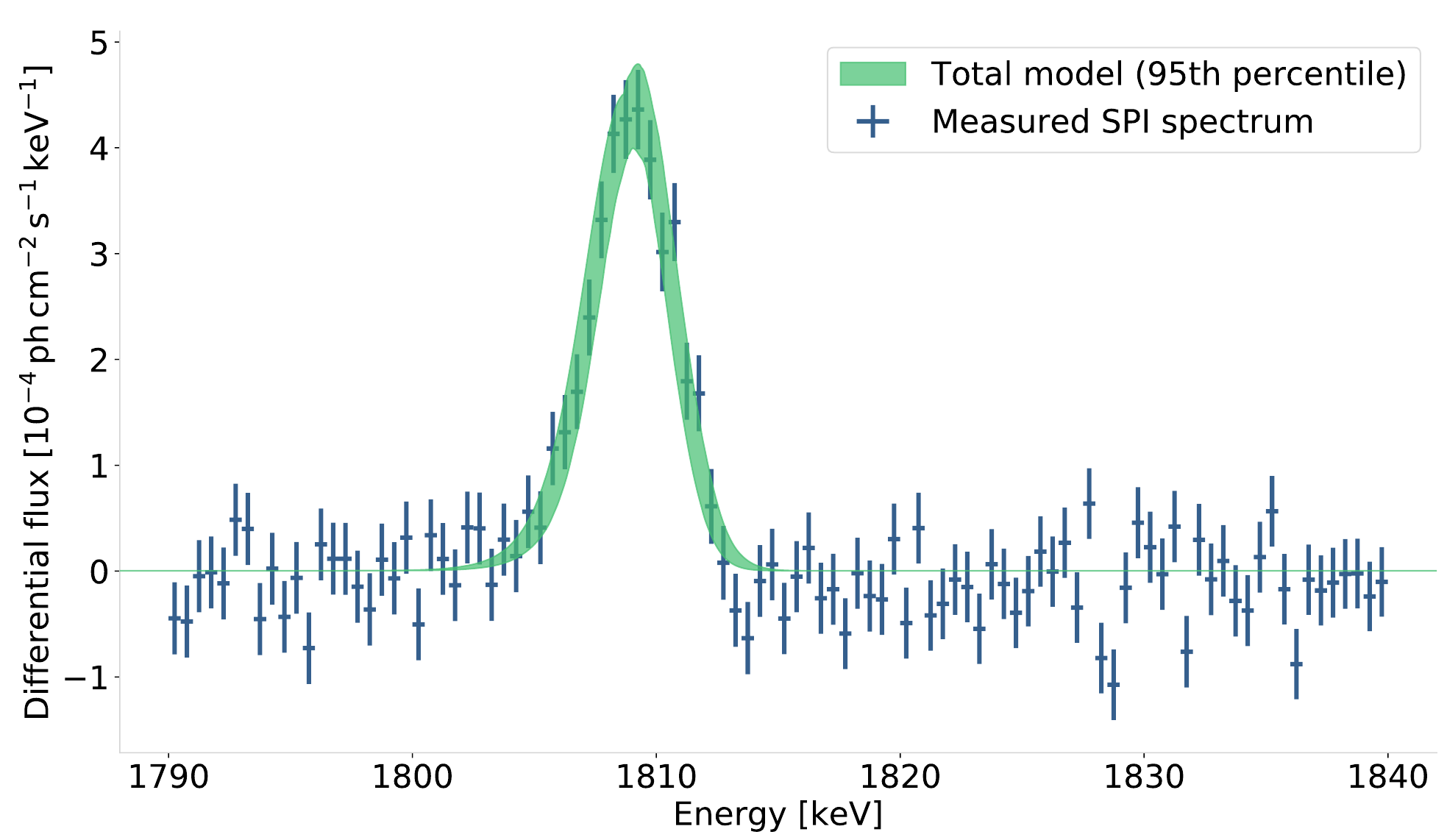}
\includegraphics[width=0.32\columnwidth,clip]{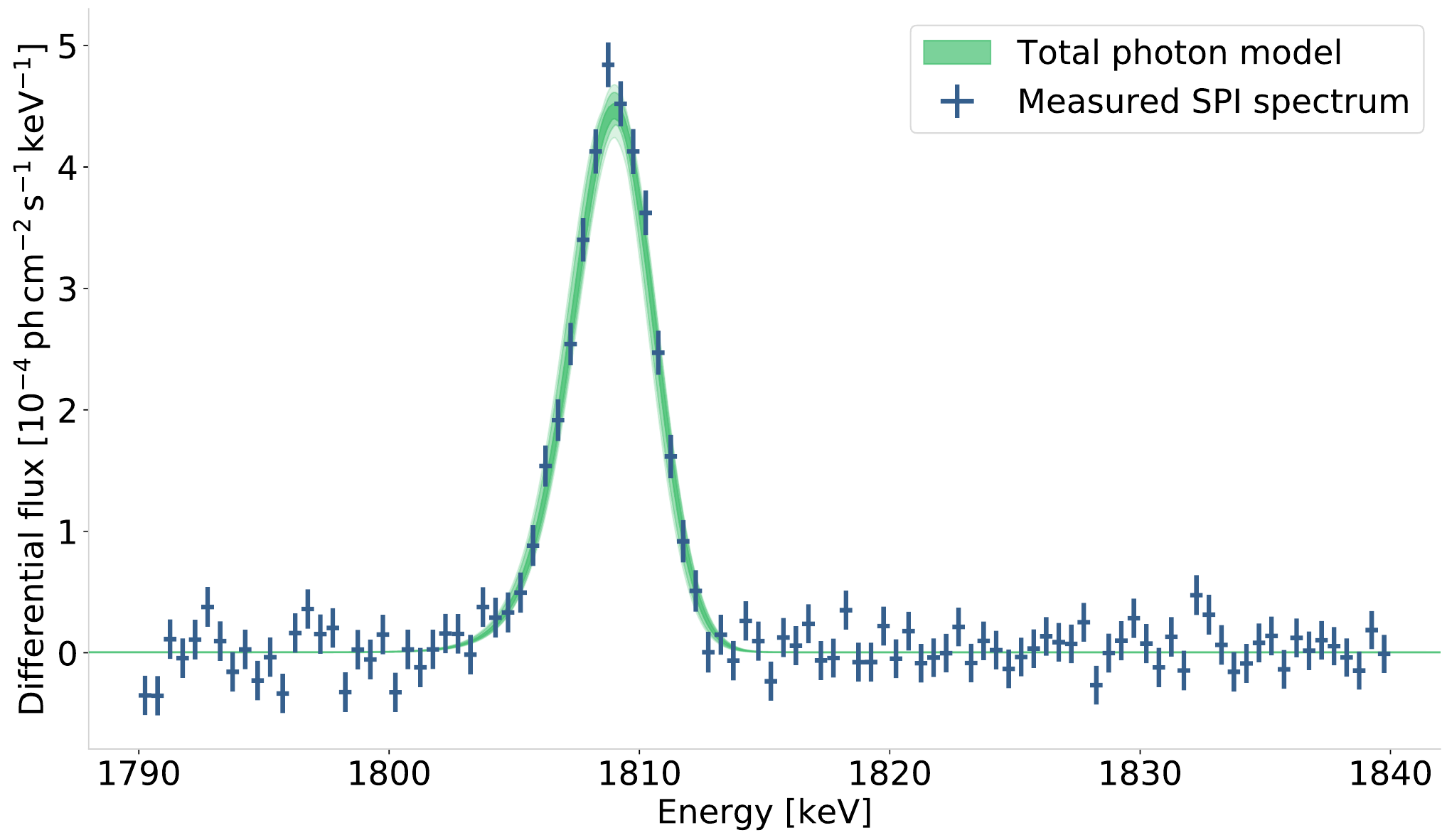}
\caption{The energy spectrum around the $^{26}$Al line, as obtained from single-hit events (left), double-hit events (middle), and combined events (right) of our observations.
Data points represent the measurement, with bin width and Poissonian statistical uncertainty indicated. The uncertainties shown in shaded green have been derived from MCMC modelling of the spectral response, and show 3$\sigma$ variations around the model.}
\label{fig:specAllsky_SE-ME}       
\end{figure} 

From 17.5 years of data and the COMPTEL-derived sky distribution of \Al emission, we obtain spectra around the \Al line at 1808.65~keV for single and for double events, separately, and combining single and double events, as shown in Figure~\ref{fig:specAllsky_SE-ME} (see \cite{Pleintinger:2023} for more details).
Intensity, line centroid, and line width are consistent within uncertainties between single-hit and double hit event results. 

\revised{The} spectral resolution of SPI \revised{up to now only has been determined} for single events as 3.17\,keV at 1.8\,MeV \citep{Diehl:2018}.
\revised{Apparently, } the spectral response appears to be less sharp for double events.
This is attributed to a bias in locations of events within Ge detectors, double events favouring events detected from interactions in outer regions, rather than closer to the central anode. Charge collection properties vary across the detector with radius from the anode, and a bias in location will plausibly produce a line shape deviating from the all-detector sample in single events (which was confirmed to be nearly Gaussian after annealings).
When decomposing the line width into contributions from instrumental resolution as measured with other instrumental lines versus line broadening of the celestial contribution, we obtain a consistent celestial broadening of  (0.46$\pm$0.3)~keV and  (0.76$\pm$0.6)~keV, respectively for single and double events, and a combined value of  (0.62$\pm$0.03)~keV.
This is consistent with the Doppler broadening from regions at different bulk velocities to the observer, as incurred from large-scale Galactic rotation and the dispersal of \Al into superbubbles, as found before \citep{Kretschmer:2013,Krause:2015}.
The line centroid, found at (1809.83$\pm$0.04)~keV from combined data, as compared to the laboratory value of 1808.65(7)~keV, shows that the integrated $^{26}$Al emission from the Galaxy appears blue-shifted by about 1.2~keV.
\revised{This is understood as the net effect of \Al bulk velocities from massive star groups across the Galaxy, as viewed from our position near the Sun \citep[see][for the variation of line centroids with galactic longitude]{Kretschmer:2013}.}

The all-sky $^{26}$Al line flux  is determined as (1.84$\pm$0.03$)\times$10$^{-3}$~ph~cm$^{-2}$s$^{-1}$. 
The flux attributed to the \emph{inner Galaxy} is $\sim$3$ \times$10$^{-4}$~ph~cm$^{-2}$s$^{-1}$, or 16\% of the total flux.
The inner Galaxy has often been  used for comparisons, and may be defined as a longitude range $\pm$30 degrees around the Galactic centre, or, slightly differently, the \emph{inner radian} (57 degrees).  This region often had been analysed as a reference towards obtaining results representative for the Galaxy as a whole, thus focusing on the brightest region and avoiding issues from low surface-brightness contributions that often are subject to systematic uncertainties from background determinations. Different latitude ranges have been used, as well, from $\pm$5 to 15 degrees.
   We note that the majority of the integrated flux is found outside the \emph{inner Galaxy}, as reported earlier \citep[see][and table 1 herein]{Pleintinger:2019}. This is in tension with simulations and expectations, which report typically 50\% of the all-sky flux coming from the inner Galaxy. Special nearby regions, such as Cygnus, Scorpius-Centaurus, and Orion, have been known before to potentially lead to distortions of such Galaxy-wide conclusions. It remains a challenge to properly account for flux contributions from nearby regions, in particular as the imaging performance of the coded mask degrades for diffuse emission that extends over large parts of the entire sky.

\begin{figure}[ht] 
\centering
\includegraphics[width=0.43\columnwidth,clip]{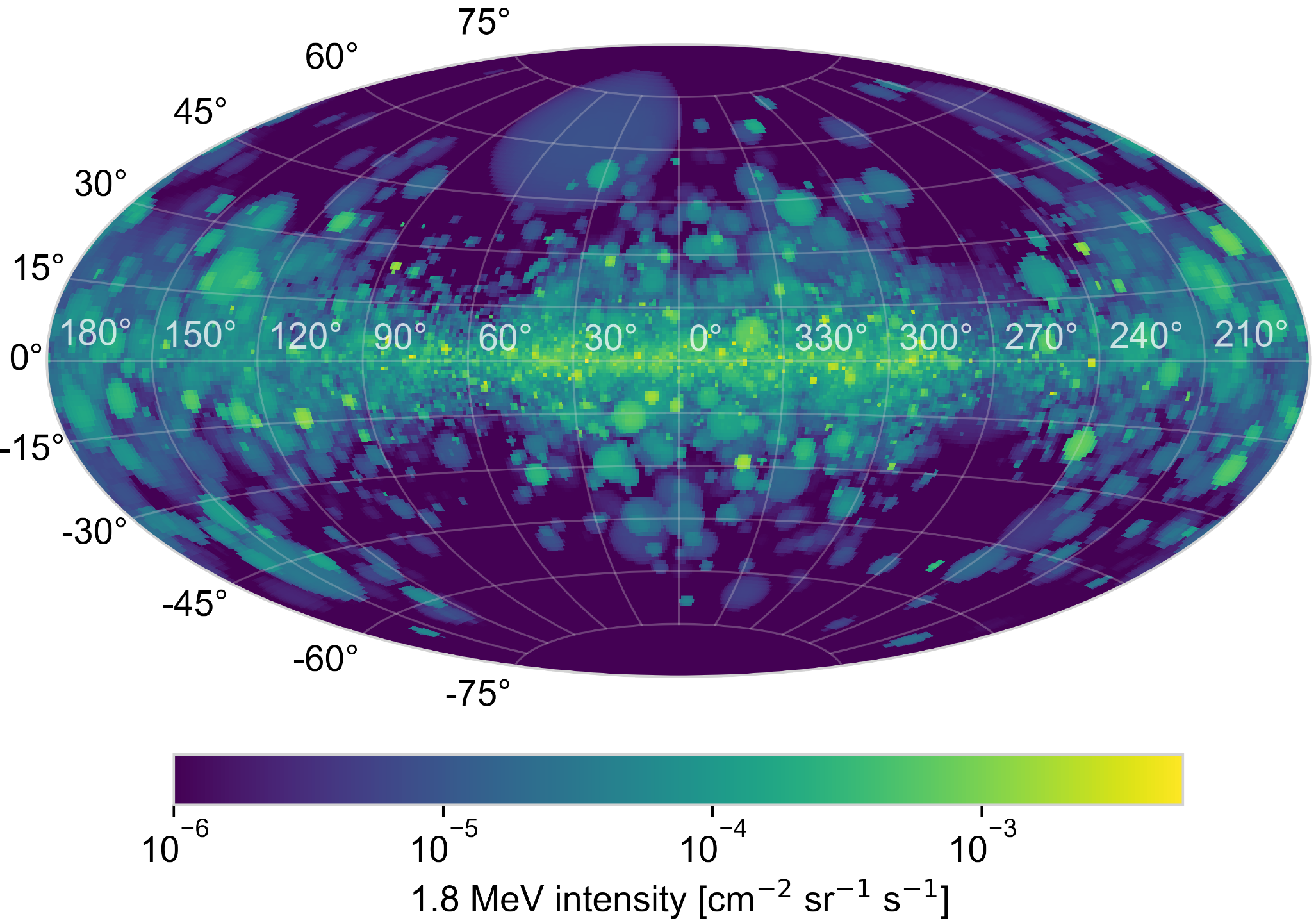}
\includegraphics[width=0.48\columnwidth,clip]{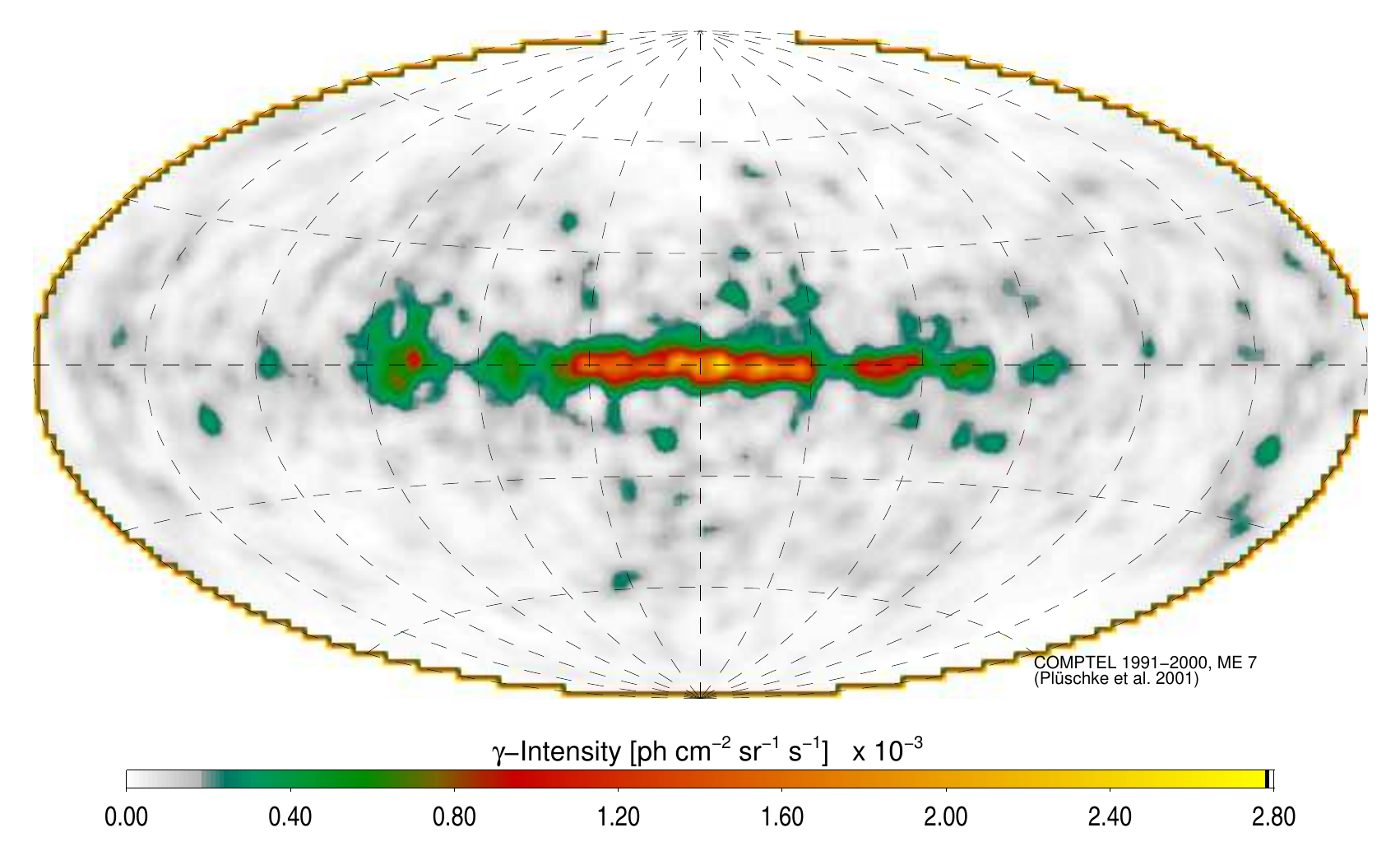}
\caption{The spatial appearance of the \Al sky as predicted from our PSYCO bottom-up massive-cluster model (\emph{left}) is similar to the COMPTEL \Al skymap (\emph{right}). Note that the model image is plotted over a larger dynamic range of intensities, compared to the COMPTEL image.}
\label{fig:imageAllsky_mod-obs}       
\end{figure} 

The bottom-up modelling as fitted to our dataset obtains a best-matching sky distribution as shown in Figure~\ref{fig:imageAllsky_mod-obs}. 
This is based on the truncated gaussian 4-armed spiral model with scale radius and width of 8.5 and 3.5 kpc, respectively, and a scale height of 700~pc.
Evidently, our superbubble modelling determines many spherical features from more-nearby clusters, which appear more regular than in reality, where superbubbles are known to be elongated, distorted, and sometimes subjective to blowouts. 
But otherwise, the essential features and large-scale appearances of this modelling and the COMPTEL \Al image also shown in Figure~\ref{fig:imageAllsky_mod-obs} for comparison are rather similar. This confirms our modelling baseline of \Al emission in our Galaxy being mainly due to massive-star groups, rather than smoother spatial distributions as would be expected if more-abundant AGB stars or novae would dominate.
We note that our bottom-up population synthesis model PSYCO predicts a substantially-lower flux value than observed: While a star formation rate  of 2-4~\Msol~y$^{-1}$ is plausible \cite{Diehl:2006d,Chomiuk:2011}, our best fit is obtained with 8~\Msol~y$^{-1}$.  This is implausibly high, and we understand this as foreground emission and sources being prominent and different in appearance than our simplified spherical superbubble modelling \citep[see][for details and discussion]{Siegert:2023}.





%
%
\bibliographystyle{woc}

%
%
%

\end{document}